\documentclass[%
 preprint,
 amsmath,amssymb,
 aps, physrev,
showkeys]{revtex4-2}

\usepackage{color}
\usepackage{graphicx}
\usepackage{dcolumn}
\usepackage{bm}

\begin{document}

\preprint{APS/123-QED}

\title{\textbf{A Classification of Hirota-Integrable Supersymmetric Bilinear KdV-Type Equations}}

\author{Laurent Delisle}
\author{Amine Jaouadi}%
 \email{Contact author: ajaouadi@ece.fr}
\affiliation{%
 LyRIDS, ECE Engineering School, OMNES Education.\\
 10 rue Sextius Michel, 75015, Paris, France
}%



\date{\today}


\begin{abstract}

We present a classification of supersymmetric bilinear KdV-type equations admitting unconstrained three-super-soliton solutions. Extending Hirota's classical three-soliton criterion to the supersymmetric setting, we derive the complete bosonic and fermionic compatibility conditions governing the existence of three-super-soliton solutions. We prove that every supersymmetric bilinear KdV-type equation possesses unconstrained one- and two-super-soliton solutions, whereas three-super-soliton solutions exist only when eight integrability conditions are satisfied. These conditions provide a supersymmetric analogue of Hirota's classical integrability criterion and naturally recover the fermionic relations previously introduced by Carstea, revealing their structural origin. As a consequence, we classify the supersymmetric extensions of Hirota bilinear KdV-type equations and show that only a subset of Hietarinta's classical classification remains valid in the unrestricted supersymmetric framework.

\end{abstract}

\maketitle


\section{\textit{Introduction}}\label{Introduction}

In the mathematical physics community, integrable nonlinear evolution equations have been largely studied in terms of their analytical structures and exact solutions. Among these solutions, we find solitons which represent localized structures of permanent form and manifest elastic interaction upon collisions with other such waves profiles. The search for such solutions have led to the development of analytical methods: inverse scattering transform \cite{Ablowitz}, Darboux and Bäcklund transformations \cite{Rogers}, and the Hirota direct method \cite{Hirota,DJZ}. The latter powerful algebraic method consists of transforming a nonlinear equation in a convenient bilinear form using a dependent variable substitution.  The obtained bilinear systems is then solved using finite expansions and modeling each solitary wave by exponential function of a certain wave number. This method allows systematic construction of multi-soliton solutions and the study of their scattering dynamics. Indeed, the two-soliton solutions showcases a constant dependent on respective wave numbers describing phase shift undergone upon interactions. The Hirota bilinear formalism remains a top research topic in the mathematical physics community \cite{zhang-liu-zhang,kumar-mohan-hbf,razzaq-zafar,ye-wang-chang-he,DJZ}. The present authors have studied such solutions in Bose-Einstein condensates \cite{celanie,delisle-symmetry}, in one-dimensional nonlinear Schrödinger equation \cite{10.1116/5.0304480} and in the financial context where the option-price is modeled by a wave function dynamically tuned by the nonlinear Schrödinger equation and exposed to a trapping potential modeling regulation \cite{Delisle-finance}. The present authors have, for the first time, developed a vector formulation of the Hirota bilinear formalism which was applied to 
the coupled modified KdV equations \cite{delisle-jaouadi} and completely integrable coupled nonlinear Schrödinger system \cite{foucher}. This new formulation allows to describe more efficiently the nonlinear coupling between each components and also one to keep the original vector representation of the systems.

The existence of multi-soliton solutions using the Hirota direct method has led to an integrability definition. We say that a system of equations written in bilinear form is Hirota integrable \cite{10.1007/BFb0113694,DJZ} if one can combine any number $N$ of one-soliton solution to construct an $N-$soliton solution, and this combination is a finite polynomial in exponential functions. It is conjectured that a bilinear equation having a three-soliton solution on the dispersion manifold is equivalent to Hirota integrability. In the 1980', Hietarinta gave a classification of such bilinear equation passing the three-soliton condition for KdV \cite{Hietarinta1}, mKdV \cite{Hietarinta2} and Sine-Gordon types \cite{Hietarinta3} equations. For complex equations \cite{Hietarinta4} such as the nonlinear Schrödinger equation, the existence of a two-soliton solution is often used as the integrability condition due to higher degree of liberty on the complex-valued dependent functions.

From a mathematical viewpoint, the three-soliton condition occupies a distinguished position in Hirota's direct method, as it provides an algebraic characterization of complete integrability for broad classes of bilinear equations. Hietarinta's classification of KdV-type bilinear equations demonstrated that this condition is sufficiently restrictive to isolate the integrable polynomial families. Extending this classification to supersymmetric bilinear equations therefore constitutes a natural and fundamental problem, where the presence of Grassmann variables introduces additional algebraic compatibility conditions absent in the classical setting.

Supersymmetric (SUSY) extensions of such integrable models arise when one seeks to enlarge the symmetry content of a nonlinear evolution equation by coupling its bosonic field to a Grassmann-odd (fermionic) superpartner. These extensions have been largely studied in terms of integrability. We note, in particular, Lax formulations \cite{popowicz-lax,krivonos-pashnev-popowicz}, Bi-hamiltonian structure \cite{Oevel}, nonlocal conservation laws \cite{andrea-restuccia-sotomayor,dargis-mathieu} and hierarchies \cite{Manin,ivanov-krivonos,McArthur,Figueroa}. Manin and Radul proposed a SUSY extension of the celebrated KdV equation \cite{Manin}, and Mathieu showed that it is invariant under a supersymmetry transformation mixing the bosonic field and its fermionic partner \cite{mathieu}. It was shown that the integrability of this extension, in the sense of possessing bosonic conserved charges in involution, holds only for isolated values of the free parameter $a=-2,1,4$ in the equation \cite{labelle-mathieu,  popowicz-lax}. This directly shows that SUSY extensions is not trivial and the study of integrability conditions is richer in this case as the fermionic contributions isn't trivial. We can thus genuinely ask how can we extend multi-soliton solutions and Hirota's integrability in a SUSY context.

The construction of super-soliton solutions of SUSY extensions of integrable models have been obtained using the SUSY extension of the Hirota bilinear formalism \cite{Carstea,Carstea1,Liu,Ghosh,Delisle,delisle-novel-hbf,ghosh-sarma-2001,delisle-n2-hbf-2017,LiuHuZhang}, Crum, Bäcklund and Darboux transformations \cite{liu-manas-darboux,liu-manas-crum,liu-manas-pfaffian,liu-darboux-1995,liu-xie,li-nimmo}. The super-Hirota bilinear method follows the same general idea as in the bosonic case. Following a dependent variable substitution, the SUSY integrable model is recast in convenient super-bilinear form. This was done by generalizing the Hirota derivative to its supersymmetric counterpart \cite{Carstea}. We also note that the Bell polynomials may come handy when constructing a bilinear representation of a soliton equation \cite{Delisle,fan-hon-2012,zhang-liu-zhang}. Soliton solutions are then constructed using finite expansion of exponential functions. Each exponential depending on a bosonic and fermionic parameter. However, the construction of multi-soliton solution of KdV-type equation is rapidly confronted to a constraint on the fermionic parameters which seems incoherent with a SUSY extension of Hirota's integrability \cite{Carstea}. The $\tau-$functions are not simple generalization of the bosonic case. For the SUSY KdV equation \cite{Carstea1,Liu} specifically, these fermionic relations have been relieved by correcting the pairwise soliton interaction constant and by dressing the fermionic part of the exponential functions. For other known SUSY extensions, the fermionic relations remain \cite{Carstea,Ghosh}.
Although these fermionic constraints successfully produce multi-super-soliton solutions, their mathematical origin has remained unclear. In particular, it is not known whether they merely arise from the specific ansatz adopted in the construction of the $\tau$-functions or whether they reflect intrinsic compatibility conditions imposed by supersymmetric Hirota integrability itself.
This naturally raises the question of whether a supersymmetric analogue of Hietarinta's classification exists and, if so, how the additional fermionic degrees of freedom modify the classical three-soliton integrability criterion. Building on Hietarinta's classification of bilinear KdV-type equations via the three-soliton condition \cite{10.1007/BFb0113694} and on Carstea's fermionic-constrained construction of SUSY solitons \cite{Carstea}, we classify SUSY extensions of bilinear KdV-type equations according to whether their three-soliton solution requires such a fermionic constraint or not.
Our approach is entirely constructive and yields explicit algebraic conditions characterizing the existence of unconstrained three-super-soliton solutions.
 Indeed, we give the three-soliton existence integrability conditions in the SUSY case. In the bosonic case, the existence of a three-soliton solution relies on one or more constraint equations \cite{10.1007/BFb0113694,Hietarinta1}. In the SUSY counterpart, we show that this existence relies on a set of eight conditions. This is highly constrainig which explains the imposed fermionic relations. We retrieve, in the bosonic limit, Hietarinta's condition for classical KdV-type equations. We also show that when one imposes the known fermionic constraints, the eight conditions reduces to two conditions: the classical bosonic condition and one further fermionic condition.
This establishes a direct correspondence between the unrestricted supersymmetric classification and the constrained constructions previously reported in the literature. It provides a systematic answer to when SUSY extension of integrable models preserves Hirota integrability outright, and when it genuinely requires additional fermionic structure.

The main contributions of this work are threefold. First, we establish the supersymmetric analogue of Hirota's classical three-soliton criterion by deriving the complete set of bosonic and fermionic compatibility conditions governing unconstrained three-super-soliton solutions. Second, these conditions naturally explain the fermionic constraints previously introduced by Carstea as structural consequences of supersymmetric integrability. Finally, they lead to a classification of supersymmetric bilinear KdV-type equations, revealing that only a subset of Hietarinta's classical classification survives in the unrestricted supersymmetric framework. This paper is constructed as follows: In section II, we recall the SUSY extension of the Hirota bilinear formalism and give a general SUSY KdV-type bilinear equation. In section III, we show that such a bilinear equation always possesses a one and two super-soliton solution, and establish the fermionic and bosonic Hirota integrability condition governing the existence of a three super-soliton solution. In section IV, we discuss in more details the integrability conditions and retrieve the fermionic parameter relationships established by Carstea from a condition on the solitary wave numbers. In section V, we give a classification of SUSY KdV-type bilinear equation passing the three-soliton integrability condition. More importantly, we retrieve Hietarinta's bosonic classification and show that in the SUSY case, only a subset of this classification passes the three super-soliton integrability condition. Last section is devoted to future perspectives and conclusions.




\section{\textit{Bilinear KdV-type equations}}\label{Bilinear}

The integrable $N=1$ supersymmetric extension of the KdV equation \cite{Manin,  mathieu} is
\begin{equation}
    \Phi_t+\Phi_{xxx}+3D^2(\Phi D\Phi)=0,
    \label{SKdV}
\end{equation}
where $\Phi$ is a fermionic superfield defined on the superspace $(x,t,\theta)$. The variables $(x,t)$ represent the usual even space-time coordinates, and $\theta$ is an odd Grassmann variable satisfying $\theta^2=0$ \cite{Cornwell:1989bx}. On the superspace, we define the covariant derivative $D=\partial_{\theta}+\theta\partial_x$ which satisfies $D^2=\partial_x$. Using a Taylor expansion around $\theta=0$, the superfield $\Phi$ may be expressed in terms of components:
\begin{equation}
    \Phi(x,t,\theta)=\zeta(x,t)+\theta u(x,t),
    \label{Superfield}
\end{equation}
where $\zeta$ and $u$ are, respectively, odd and even functions of $x$ and $t$. Substituting the superfield extension into the SUSY KdV equation (\ref{SKdV}) yield a system of two equation (one even, one odd):
\begin{eqnarray}
    u_t+u_{xxx}+6uu_x-3\zeta\zeta_{xx}&=&0,\\
    \zeta_t+\zeta_{xxx}+3(\zeta u)_x&=&0.
\end{eqnarray}
In the bosonic limit, where $\zeta\longrightarrow0$, the system reduces to the classical KdV equation. Soliton solutions of equation (\ref{SKdV}) have been constructed using the bilinear transformation \cite{Carstea,Carstea1,McArthur}
\begin{equation}
    \Phi=2 D^3\log\tau,
    \label{KdVSubs}
\end{equation}
where $\tau=\tau(x,t,\theta)$ is a bosonic superfield. This substitution allows a supersymmetric Hirota bilinear representation of the KdV equation \cite{Carstea,McArthur} has
\begin{equation}
   \mathcal{S}\mathcal{P}(\mathcal{D}_x,\mathcal{D}_t)(\tau\cdot\tau)=0,
   \label{BiliSKdV}
\end{equation}
where $\mathcal{P}(X,T)=X^3+T$. The super-Hirota derivative \cite{Carstea} is defined as
\begin{equation}
    \mathcal{S}\mathcal{D}_{z}^n(\mu\cdot \nu)=(D_{1}-D_2)(\partial_{z_1}-\partial_{z_2})^n\mu(z_1;\theta_1)\nu(z_2;\theta_2)\vert_{z_1=z_2=z}^{D_1=D_2=D}.
\end{equation}
The bilinear equation (\ref{BiliSKdV}) was shown to possess $N-$soliton solution. However, the first construction \cite{Carstea} imposes linear relations between the fermionic parameters involve in the construction of the solitons. This contradicts the classical definition of Hirota integrability \cite{10.1007/BFb0113694} where a $N-$soliton is constructed from a finite polynomial combination of one-soliton solution without any constraints on the parameters a part from the usual dispersion relations. The fermionic linear relation was relaxed \cite{Carstea1} by assuming that the interaction parameters depends on the fermionic parameters and by a dressing method on the exponential modeling pairwise soliton interaction. This shows that the $\tau-$functions for the construction of $N-$soliton solution of SUSY extensions of integrable models are not merely trivial and need to be corrected by fermionic contributions.
The central problem is therefore to determine which polynomials $\mathcal{P}$ give rise to supersymmetric bilinear equations admitting unconstrained multi-super-soliton solutions. This transforms the question of supersymmetric integrability into an algebraic classification problem.
In this paper, we propose to classify SUSY bilinear equations of KdV-type (\ref{BiliSKdV}) passing the three-soliton integrability test. We assume that the polynomial $\mathcal{P}$ is odd \textit{i.e.} $\mathcal{P}(-X,-T)=-\mathcal{P}(X,T)$. 

For all such polynomials $\mathcal{P}$, we prove that a one- and a two-soliton solution can always be constructed assuming only the dispersion relations to be satisfied. For the three-soliton soliton, we show that the dispersion relations alone are not sufficient for its existence and to ensure parameter freedom. The polynomial $\mathcal{P}$ should also satisfy a set of conditions, that we will call later on the Hirota integrability conditions. The idea is to retrieve polynomial $\mathcal{P}$ that satisfy these conditions on and only on the dispersion manifold \cite{10.1007/BFb0113694}.
The remainder of the paper is devoted to deriving these compatibility conditions explicitly and determining all polynomial families satisfying them. This ultimately yields the supersymmetric counterpart of Hietarinta's classification for bilinear KdV-type equations.

\section{Soliton Solution}

In this section, we prove that supersymmetric bilinear equations of the form (\ref{SKdV}) possess one and two-soliton solution. For the three soliton solution, additional integrability conditions are needed to classify polynomials $\mathcal{P}$.
The construction proceeds incrementally by considering one-, two-, and three-super-soliton solutions. While the first two cases exist for arbitrary odd polynomials satisfying the corresponding dispersion relations, the three-super-soliton solution reveals the additional algebraic compatibility conditions that characterize supersymmetric Hirota integrability.

\subsection{One-soliton solution}
For the one-soliton solution, we assume the following form for that $\tau-$function:
\begin{equation}
    \tau=1+e^{\eta+\theta\xi},
\end{equation}
where $\eta=\kappa x+\omega t+\varphi$ and $\xi$ an odd constant. We assume here and further on that $\kappa$, $\omega$ and $\varphi$ are real numbers. Introducing the $\tau-$function ansatz in the bilinear equation (\ref{BiliSKdV}), we get the dispersion relation
\begin{equation}
    \mathcal{P}(\vec{\eta})=0,
\end{equation}
where $\eta=\kappa x+\omega t+\varphi=\vec{\eta}\cdot\vec{x}+\varphi$ with $\vec{\eta}=(\kappa,\omega)$ and $\vec{x}=(x,t)$. In particular, for the SUSY KdV equation (\ref{SKdV}), we have $\mathcal{P}(X,T)=X^3+T$ and we get the well-known dispersion relation $\omega=-\kappa^3$.

From the dependent variable substitution (\ref{KdVSubs}), we get the components of the superfied $\Phi$ given in Eq.(\ref{Superfield}):
\begin{equation}
    \zeta=\frac{\xi\kappa}{2}\mbox{sech}^2\left(\frac{\eta}{2}\right)\quad \mbox{and}\quad u=\frac{\kappa^2}{2}\mbox{sech}^2\left(\frac{\eta}{2}\right),
\end{equation}
which represent classical soliton lump profiles.

The existence of the one-super-soliton solution depends solely on the dispersion relation (Eq.(9)) and is therefore independent of the particular choice of the polynomial $\mathcal{P}$. This universality provides the starting point for the subsequent integrability analysis.

\subsection{Two-soliton solution}

For the two-soliton solution, the SUSY extension of the $\tau-$function is not as straightforward as for the classical case \cite{Carstea1}. Indeed, if we assume a trivial extension of the classical case to the supersymmetric one, the two-soliton existence relies on a linear relations between the fermionic parameters \cite{Carstea}. This contradicts the essence of Hirota's integrability \cite{10.1007/BFb0113694} as it relies on the freedom of the parameters a part from the usual dispersion relations. Therefore, as in \cite{Carstea1}, we use a dressing method to relieve the two-soliton solution of this fermionic linear relation. We thus assume the $\tau-$function to have the form
\begin{equation}
    \tau=1+e^{\eta_1+\theta\xi_1}+e^{\eta_2+\theta\xi_2}+A_{12}e^{\eta_1+\eta_2+\theta\xi_{12}},\label{tau2soliton}
\end{equation}
where $\eta_{m}=\kappa_m x+\omega_mt+\varphi_m$, $A_{12}$ is an even constant and $\xi_1, \xi_2, \xi_{12}$ are odd parameters. Introducing this $\tau-$function in the bilinear equation (\ref{BiliSKdV}), we get the dispersion relations $\mathcal{P}(\vec{\eta}_{m})=0$ for $m=1,2$. The unknowns $A_{12}$ and $\xi_{12}$ are explicitly given as
\begin{equation}
    A_{12}=B_{12}\left(1-2\frac{\xi_1\xi_2}{\kappa_1-\kappa_2}\right)=A_{21}\quad \mbox{and}\quad
    \xi_{12}=\alpha_{12}\xi_1+\alpha_{21}\xi_2=\xi_{21},
\end{equation}
where
\begin{equation}
    \alpha_{12}=-\alpha_{21}=\frac{\kappa_1+\kappa_2}{\kappa_1-\kappa_2}\quad\mbox{and}\quad B_{12}=-\frac{1}{\alpha_{12}}\frac{\mathcal{P}(\vec{\eta}_1-\vec{\eta}_2)}{\mathcal{P}(\vec{\eta}_1+\vec{\eta}_2)}=B_{21}.
\end{equation}
In the fermionic limit $\xi_1,\xi_2\longrightarrow0$, the constant $A_{12}$ reduces to the classical pairwise soliton interaction parameter $B_{12}$ used to describe phase shift dislocations \cite{Hirota,10.1116/5.0304480,Hietarinta1}. The above constitute an unconstrained two-soliton solution on the dispersion manifold $\mathcal{M}_2=\{(\vec{\eta}_1,\vec{\eta}_2) : \mathcal{P}(\vec{\eta}_1)=\mathcal{P}(\vec{\eta}_2)=0\}$, the unknown parameter $A_{12}$ is directly linked to the polynomial $\mathcal{P}$. We can thus assert that a supersymmetric bilinear KdV-type equation always possesses  a two-soliton solution.
This result shows that, as in the classical Hirota formalism, the existence of unconstrained two-super-soliton solutions is not sufficiently restrictive to characterize integrability. The first nontrivial obstruction arises only at the level of the three-super-soliton solution.
This is coherent to the result obtained by Hietarinta \cite{Hietarinta1}. As we will see in the next section, a three-soliton solution is not automatic and its existence will require integrability conditions.

Using the $\tau-$function representation discussed in \cite{LiuHuZhang}, the $\tau-$function, given in equation (\ref{tau2soliton}), may be equivalently expressed as
\begin{equation}
    \tau=1+e^{\check{\eta}_1}+e^{\check{\eta}_2}+\check{A}_{12}e^{\check{\eta}_1+\check{\eta}_2},\label{tauRep}
\end{equation}
where $\check{\eta}_p=\eta_p+\theta\xi_p$ for $p=1,2$. This representation is identical to the $\tau-$function representation in the non-supersymmetric KdV equation. The interaction coefficient $\check{A}_{12}$ is explicitly given by
\begin{equation}
    \check{A}_{12}=B_{12}\left(1-2\frac{\xi_1\xi_2}{\kappa_1+\kappa_2}+2\theta\frac{(\kappa_2\xi_1-\kappa_1\xi_2)}{\kappa_1+\kappa_2}\right).\label{checkA12}
\end{equation}
This convenient form shows that $\check{A}_{12}$ reduces to $B_{12}$ if and only if $\kappa_2\xi_1=\kappa_1\xi_2$. This linear relation between the fermionic parameters $\xi_1$ and $\xi_2$ is identical to the one obtained by Carstea for the existence of two- and multi-super-soliton solution of a certain class of SUSY extensions of integrable models \cite{Carstea}. However, this linear constraint is not necessary for SUSY bilinear KdV-type equations of the form (\ref{BiliSKdV}). In this paper, we choose the $\tau-$function representation (\ref{tau2soliton}) rather then (\ref{tauRep}). This choice is made for several reasons: to stay faithful to the original representation of the SUSY $\tau-$function \cite{Carstea1} and to ensure that the Grassmann variable $\theta$ is exclusively present in the exponential functions in order to facilitate symbolic calculations using the Hirota super-derivative.

\begin{figure*}[ht]
\centering
\includegraphics[width=1.0\textwidth]{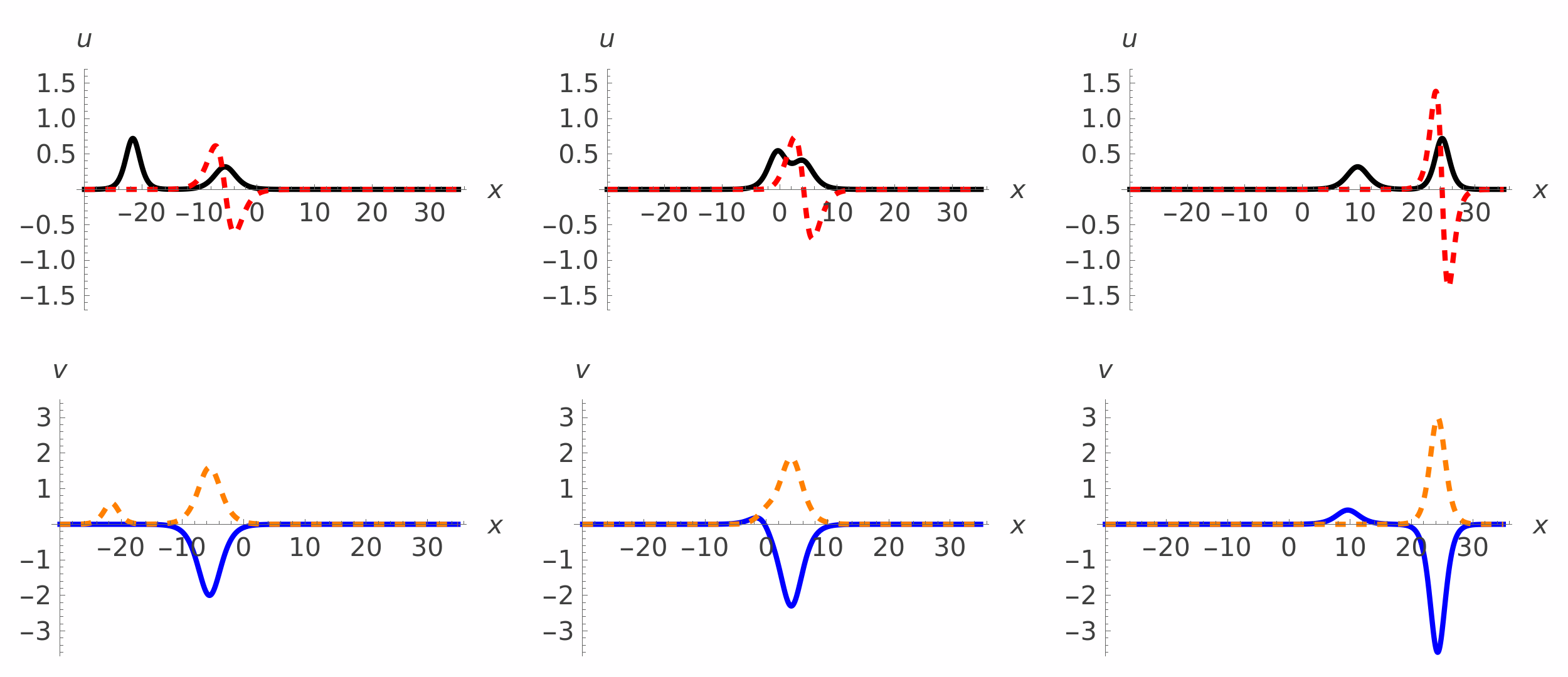}
\caption{The two-super-soliton solution for the SUSY KdV equation (\ref{SKdV}). The parameter choices are: $\kappa_1=0.8$, $\kappa_2=1.2$ and $\varphi_1=\varphi_2=0$. The left, middle and right columns are for, respectively, time $t=-15,0,15$. The upper row represents the curves $u_0$ (full black curve) and $u_{12}$ (dashed red curve). The bottom row represents the curves $v_1$ (full blue curve) and $v_2$ (dashed orange curve).}
\label{Susy2Soliton}
\end{figure*}

To give a representation of the two-soliton solution, we write the components $u$ and $\zeta$ of the superfield $\Phi$ given in Eq.(\ref{Superfield}) as
\begin{equation}
    u=u_0+\xi_1\xi_2u_{12}\quad\mbox{and}\quad \zeta=\xi_1v_1+\xi_2v_2.
\end{equation}
We find, explicitly,
\begin{equation}
    v_m=2\left(\frac{c_m}{b_0}\right)_x,\quad u_0=2(\log b_0)_{xx}\quad \mbox{and}\quad u_{12}=2\left(\frac{b_{12}}{b_0}\right)_{xx},
\end{equation}
for $m=1,2$ and where
\begin{eqnarray}
    b_0&=&1+e^{\eta_1}+e^{\eta_2}+B_{12}e^{\eta_1+\eta_2},\quad
    b_{12}=-2\frac{B_{12}}{\kappa_1-\kappa_2}e^{\eta_1+\eta_2},\\
    c_1&=&e^{\eta_1}+B_{12}\alpha_{12}e^{\eta_1+\eta_2},\quad
    c_2=e^{\eta_2}+B_{12}\alpha_{21}e^{\eta_1+\eta_2}.
\end{eqnarray}
In Figure~\ref{Susy2Soliton}, we show the bosonic components $u_0, u_{12}$ and the fermionic components $v_1, v_2$ for the KdV polynomial $\mathcal{P}(X,T)=X^3+T$. The function $u_0$ (full black curve) represents a classic two-soliton KdV-type solution. The greater amplitude soliton travels at higher velocity and comes into interaction with the lower amplitude soliton. They interact elastically, manifesting an apparent phase shift upon collision. The correction component $u_{12}$ (dashed red curve) moves locally like a pulse-anti-pulse wave and has greater amplitude upon interaction between the two-soliton wave profiles depicted by $u_0$. The amplitude stabilizes. The fermionic components $v_1$ (full blue curve) and $v_2$ (dashed orange curve) have asymmetric dynamics. Indeed, the component $v_1$ looks like a one-soliton solution and as time evolves develop into a sort of bright-dark soliton. The component $v_2$ looks like a classical KdV two soliton soliton, but the two traveling waves seems to merge to form a higher amplitude unique solitary wave. We can observe nevertheless that susy extensions of integrable systems allows richer dynamics and fermionic correction are manifest.

Furthermore, our formalism allows one to generate the two-soliton solution for any polynomial $\mathcal{P}$ from an original integrable or non-integrable system. Confirming that many systems possess two-soliton solution despite non-integrability in the classical IST sense \cite{Ablowitz}.

\subsection{Three-soliton solution}
The two-soliton solution existence is not proof of Hirota's integrability \cite{10.1007/BFb0113694,Hietarinta1,Hietarinta2,Hietarinta3}. 
Consequently, the determination of the three-super-soliton solution constitutes the fundamental step of the classification, since it is precisely at this order that the supersymmetric compatibility conditions emerge.
Indeed, it was shown that some non-integrable systems possess one and two-soliton solution. However, the existence of an unconstrained three-soliton solution is viewed in the classical case as an equivalent definition of integrability as in the sense of a Lax pair.

In this section, we propose to find the SUSY Hirota's integrability condition for the existence of a three-soliton solution which should be verified only on the three-soliton dispersion manifold  $\mathcal{M}_2=\{(\vec{\eta}_1,\vec{\eta}_2,\vec{\eta}_3) : \mathcal{P}(\vec{\eta}_1)=\mathcal{P}(\vec{\eta}_2)=\mathcal{P}(\vec{\eta}_3)=0\}$. 

We assume the following ansatz for the $\tau-$function \cite{Carstea1}:
\begin{eqnarray}
    \tau&=&1+e^{\eta_1+\theta\xi_1}+e^{\eta_2+\theta \xi_2}+e^{\eta_3+\theta\xi_3}+A_{12}e^{\eta_1+\eta_2+\theta\xi_{12}}
    +A_{13}e^{\eta_1+\eta_3+\theta\xi_{13}}+A_{23}e^{\eta_2+\eta_3+\theta\xi_{23}}\nonumber\\&+&A_{123}e^{\eta_1+\eta_2+\eta_3+\theta\xi_{123}},\label{tau3Rep}
\end{eqnarray}
where $\eta_p=\kappa_px+\omega_pt+\varphi_p$, the $A'$s are even constants and $\xi_{pq}, \xi_{123}$ are odd constants to be determined. As for the two-soliton solution, we get
\begin{equation}
    A_{pq}=B_{pq}\left(1-2\frac{\xi_p\xi_q}{\kappa_p-\kappa_q}\right)=A_{qp}\quad \mbox{and}\quad
    \xi_{pq}=\alpha_{pq}\xi_p+\alpha_{qp}\xi_q=\xi_{qp},
\end{equation}
where
\begin{equation}
    \alpha_{pq}=-\alpha_{qp}=\frac{\kappa_p+\kappa_q}{\kappa_p-\kappa_q}\quad \mbox{and}\quad
    B_{pq}=-\frac{1}{\alpha_{pq}}\frac{\mathcal{P}(\vec{\eta}_p-\vec{\eta}_q)}{\mathcal{P}(\vec{\eta}_p+\vec{\eta}_q)}=B_{qp}.\label{BPQ}
\end{equation}

The real test starts with the determination of the unknown parameters $A_{123}$ and $\xi_{123}$. We find
\begin{equation}
    \xi_{123}=\alpha_{12}\alpha_{13}\xi_1+\alpha_{21}\alpha_{23}\xi_2+\alpha_{31}\alpha_{32}\xi_3.
\end{equation}
For $A_{123}$, we get the expression
\begin{equation}
    A_{123}=B_{12}B_{13}B_{23}\left(1-2\frac{\alpha_{13}\alpha_{23}}{\kappa_1-\kappa_2}\xi_1\xi_2\right)\left(1-2\frac{\alpha_{12}\alpha_{32}}{\kappa_1-\kappa_3}\xi_1\xi_3\right)\left(1-2\frac{\alpha_{21}\alpha_{31}}{\kappa_2-\kappa_3}\xi_2\xi_3\right).
\end{equation}
We observe that the $\tau-$function is invariant under the action of the permutation group $S_3=<(12),(123)>$. Indeed, for all permutations $\sigma\in S_3$, we define the action $\sigma\cdot(\vec{\eta},\vec{\xi})=\sigma\cdot(\eta_1,\eta_2,\eta_3,\xi_1,\xi_2,\xi_3)=(\eta_{\sigma(1)},\eta_{\sigma(2)},\eta_{\sigma(3)},\xi_{\sigma(1)},\xi_{\sigma(2)},\xi_{\sigma(3)})$. It is direct to show that for all $\sigma\in S_3$, we have $\tau(\sigma\cdot(\vec{\eta},\vec{\xi}))=\tau(\vec{\eta},\vec{\xi}). $ This shows that the $\tau-$function is invariant under the $S_3$ action, as expected, since this merely corresponds to the relabelling of the solitons. 

We are left to verify one last equation which should constitute the Hirota integrability condition. We obtain explicitly the two $S_3$-invariant Hirota integrability conditions:
\begin{eqnarray}
    A_{23}(\xi_1-\xi_{23})\mathcal{P}(\vec{\eta}_1-\vec{\eta}_2-\vec{\eta}_3)+A_{13}(\xi_2-\xi_{13})\mathcal{P}(\vec{\eta}_2-\vec{\eta}_1-\vec{\eta}_3)
    \nonumber\\+A_{12}(\xi_3-\xi_{12})\mathcal{P}(\vec{\eta}_3-\vec{\eta}_1-\vec{\eta}_2)+A_{123}\xi_{123}\mathcal{P}(\vec{\eta}_1+\vec{\eta}_2+\vec{\eta}_3)=0\label{Condition1}
\end{eqnarray}
and
\begin{eqnarray}
    A_{23}(\kappa_1-\kappa_2-\kappa_3-2\xi_1\xi_{23})\mathcal{P}(\vec{\eta}_1-\vec{\eta}_2-\vec{\eta}_3)
    \nonumber\\+A_{13}(\kappa_2-\kappa_1-\kappa_3-2\xi_2\xi_{13})\mathcal{P}(\vec{\eta}_2-\vec{\eta}_1-\vec{\eta}_3)\nonumber\\
    +A_{12}(\kappa_3-\kappa_1-\kappa_2-2\xi_3\xi_{12})\mathcal{P}(\vec{\eta}_3-\vec{\eta}_1-\vec{\eta}_2)
    \nonumber\\+A_{123}(\kappa_1+\kappa_2+\kappa_3)\mathcal{P}(\vec{\eta}_1+\vec{\eta}_2+\vec{\eta}_3)=0\label{Condition2}
\end{eqnarray}

Eq.(\ref{Condition1}) and Eq.(\ref{Condition2}) constitute the supersymmetric analogue of Hirota's classical three-soliton condition. Unlike the purely bosonic case, the presence of Grassmann variables naturally separates the compatibility conditions into even and odd sectors, whose simultaneous satisfaction is necessary for the existence of unconstrained three-super-soliton solutions.

The explicit decomposition of these supersymmetric compatibility conditions into independent bosonic and fermionic components provides the basis for the classification developed in the next section.

\section{Hirota integrability conditions}


The compatibility conditions (\ref{Condition1}) and (\ref{Condition2}) obtained in the previous section constitute the supersymmetric analogue of Hirota's three-soliton condition. Their algebraic structure is, however, not immediately transparent, since they simultaneously involve bosonic and fermionic contributions. The purpose of this section is to decompose these conditions into independent Grassmann components, determine the resulting scalar compatibility equations, and identify the algebraic relations between them. This analysis will ultimately explain the origin of the fermionic constraints appearing in supersymmetric multi-soliton constructions.
Condition (\ref{Condition1}) is odd, it can be decomposed into four conditions for the terms multiplying $\xi_1$, $\xi_2$, $\xi_3$ and $\xi_1\xi_2\xi_3$. Indeed, we may write condition (\ref{Condition1}) as
\begin{equation}
    \xi_1\Delta_1+\xi_2\Delta_2+\xi_3\Delta_3+\xi_{1}\xi_2\xi_3\Delta=0,\label{FermionicEq}
\end{equation}
which is equivalent, for totally free fermionic parameters $\xi_1$, $\xi_2$ and $\xi_3$, to $\Delta_1=\Delta_2=\Delta_3=\Delta=0$. The action of the $S_3$ generator on each component equations may be calculated. Explicitly, we have $(12)\cdot(\Delta_1,\Delta_2,\Delta_3,\Delta)=(\Delta_2,\Delta_1,\Delta_3,-\Delta)$ and $(123)\cdot(\Delta_1,\Delta_2,\Delta_3,\Delta)=(\Delta_2,\Delta_3,\Delta_1,\Delta)$. The system is thus $S_3-$invariant.  We get, explicitly
\begin{eqnarray}
    \Delta_1=B_{12}B_{13}B_{23}\alpha_{12}\alpha_{13}\mathcal{P}(\vec{\eta}_1+\vec{\eta}_2+\vec{\eta}_3)+B_{23}\mathcal{P}(\vec{\eta}_1-\vec{\eta}_2-\vec{\eta}_3)\nonumber\\-B_{13}\alpha_{13}\mathcal{P}(\vec{\eta}_2-\vec{\eta}_1-\vec{\eta}_3)-B_{12}\alpha_{12}\mathcal{P}(\vec{\eta}_3-\vec{\eta}_1-\vec{\eta}_2)&=&0,\label{F1}\\
    \Delta_2=B_{12}B_{13}B_{23}\alpha_{21}\alpha_{23}\mathcal{P}(\vec{\eta}_1+\vec{\eta}_2+\vec{\eta}_3)+B_{13}\mathcal{P}(\vec{\eta}_2-\vec{\eta}_1-\vec{\eta}_3)\nonumber\\-B_{23}\alpha_{23}\mathcal{P}(\vec{\eta}_1-\vec{\eta}_2-\vec{\eta}_3)-B_{12}\alpha_{21}\mathcal{P}(\vec{\eta}_3-\vec{\eta}_1-\vec{\eta}_2)&=&0,\label{F2}\\
    \Delta_3=B_{12}B_{13}B_{23}\alpha_{31}\alpha_{32}\mathcal{P}(\vec{\eta}_1+\vec{\eta}_2+\vec{\eta}_3)+B_{12}\mathcal{P}(\vec{\eta}_3-\vec{\eta}_1-\vec{\eta}_2)\nonumber\\-B_{23}\alpha_{32}\mathcal{P}(\vec{\eta}_1-\vec{\eta}_2-\vec{\eta}_3)-B_{13}\alpha_{31}\mathcal{P}(\vec{\eta}_2-\vec{\eta}_1-\vec{\eta}_3)&=&0\label{F3}
\end{eqnarray}
and the last equation associated to the term $\xi_1\xi_2\xi_3$:
\begin{eqnarray}
    \Delta=B_{12}B_{13}B_{23}\mathcal{P}(\vec{\eta}_1+\vec{\eta}_2+\vec{\eta}_3)\left(\frac{(\alpha_{12}\alpha_{13})^2}{\kappa_2-\kappa_3}-\frac{(\alpha_{12}\alpha_{23})^2}{\kappa_1-\kappa_3}+\frac{(\alpha_{13}\alpha_{23})^2}{\kappa_1-\kappa_2}\right)\nonumber\\
    +\frac{B_{23}}{\kappa_2-\kappa_3}\mathcal{P}(\vec{\eta}_1-\vec{\eta}_2-\vec{\eta}_3)-\frac{B_{13}}{\kappa_1-\kappa_3}\mathcal{P}(\vec{\eta}_2-\vec{\eta}_1-\vec{\eta}_3)
    +\frac{B_{12}}{\kappa_1-\kappa_2}\mathcal{P}(\vec{\eta}_3-\vec{\eta}_1-\vec{\eta}_2)=0\label{F4}.
\end{eqnarray}

Let us now consider the second Hirota integrability condition (\ref{Condition2}). This condition is even and will be decomposed into four sub-conditions for $1$, $\xi_1\xi_2$, $\xi_1\xi_3$ and $\xi_2\xi_3$. The purely bosonic condition associated to $1$ is given by
\begin{eqnarray}
    \mathcal{E}=B_{12}B_{13}B_{23}(\kappa_1+\kappa_2+\kappa_3)\mathcal{P}(\vec{\eta}_1+\vec{\eta}_2+\vec{\eta}_3)
    +B_{23}(\kappa_1-\kappa_2-\kappa_3)\mathcal{P}(\vec{\eta}_1-\vec{\eta}_2-\vec{\eta}_3)\nonumber\\+B_{13}(\kappa_2-\kappa_1-\kappa_3)\mathcal{P}(\vec{\eta}_2-\vec{\eta}_1-\vec{\eta}_3)+
    B_{12}(\kappa_3-\kappa_1-\kappa_2)\mathcal{P}(\vec{\eta}_3-\vec{\eta}_1-\vec{\eta}_2)=0.\label{BosoTest}
\end{eqnarray}
The above equation is equivalent to the classical Hirota's three-soliton condition for bilinear KdV equations. In the SUSY classification, we will start by identifying polynomials $\mathcal{P}$ this equation ($\mathcal{E}=0$) before moving on to the other conditions. For the three other equations, we get
\begin{eqnarray}
    &\phantom{=}&B_{12}B_{13}B_{23}\alpha_{13}\alpha_{23}\left(\frac{\kappa_1+\kappa_2+\kappa_3}{\kappa_1-\kappa_2}\right)\mathcal{P}(\vec{\eta}_1+\vec{\eta}_2+\vec{\eta}_3)
    +B_{12}\left(\frac{\kappa_3-\kappa_1-\kappa_2}{\kappa_1-\kappa_2}\right)\mathcal{P}(\vec{\eta}_3-\vec{\eta}_1-\vec{\eta}_2)\nonumber\\
    &+&\alpha_{23}B_{23}\mathcal{P}(\vec{\eta}_1-\vec{\eta}_2-\vec{\eta}_3)-\alpha_{13}B_{13}\mathcal{P}(\vec{\eta}_2-\vec{\eta}_1-\vec{\eta}_3)=0\\
    &\phantom{=}&B_{12}B_{13}B_{23}\alpha_{12}\alpha_{32}\left(\frac{\kappa_1+\kappa_2+\kappa_3}{\kappa_1-\kappa_3}\right)\mathcal{P}(\vec{\eta}_1+\vec{\eta}_2+\vec{\eta}_3)
    +B_{13}\left(\frac{\kappa_2-\kappa_1-\kappa_3}{\kappa_1-\kappa_3}\right)\mathcal{P}(\vec{\eta}_2-\vec{\eta}_1-\vec{\eta}_3)\nonumber\\
    &-&\alpha_{23}B_{23}\mathcal{P}(\vec{\eta}_1-\vec{\eta}_2-\vec{\eta}_3)-\alpha_{12}B_{12}\mathcal{P}(\vec{\eta}_3-\vec{\eta}_1-\vec{\eta}_2)=0\\
    &\phantom{=}&B_{12}B_{13}B_{23}\alpha_{21}\alpha_{31}\left(\frac{\kappa_1+\kappa_2+\kappa_3}{\kappa_2-\kappa_3}\right)\mathcal{P}(\vec{\eta}_1+\vec{\eta}_2+\vec{\eta}_3)
    +B_{23}\left(\frac{\kappa_1-\kappa_2-\kappa_3}{\kappa_2-\kappa_3}\right)\mathcal{P}(\vec{\eta}_1-\vec{\eta}_2-\vec{\eta}_3)\nonumber\\
    &+&\alpha_{31}B_{13}\mathcal{P}(\vec{\eta}_2-\vec{\eta}_1-\vec{\eta}_3)-\alpha_{21}B_{12}\mathcal{P}(\vec{\eta}_3-\vec{\eta}_1-\vec{\eta}_2)=0
\end{eqnarray}
All the above conditions are $S_3-$invariant as it should be.

An important observation linking the purely bosonic constraint $\mathcal{E}=0$ and the fermionic components $\Delta_1,\Delta_2$ and $\Delta_3$ can be made:  if $\mathcal{E}=0$, then the equation
\begin{equation}
    \Delta_1+\lambda\Delta_2+\gamma\Delta_3=0
\end{equation}
if and only if
\begin{eqnarray}
    \kappa_2\kappa_3(\kappa_2-\kappa_3)+\lambda\kappa_1\kappa_3(\kappa_3-\kappa_1)+\gamma\kappa_1\kappa_2(\kappa_1-\kappa_2)=0 \quad \mbox{or}\label{kinematic}\\
    (\kappa_2+\kappa_3)B_{23}\mathcal{P}(\vec{\eta}_1-\vec{\eta_2}-\vec{\eta_3})+(\kappa_1+\kappa_3)B_{13}\mathcal{P}(\vec{\eta}_2-\vec{\eta_1}-\vec{\eta_3})\nonumber\\+(\kappa_1+\kappa_2)B_{12}\mathcal{P}(\vec{\eta}_3-\vec{\eta_1}-\vec{\eta_2})=0\label{additional}.
\end{eqnarray}
The second equation is constraining on the polynomial $\mathcal{P}$. It is satisfied for polynomial of the type $\mathcal{P}=a X^3+bX+T$, but not for polynomial of the type $\mathcal{P}=a X^5+b X^3+cX+T$ for $a\neq 0$. This latter polynomial does however satisfy the purely bosonic constraint $\mathcal{E}=0$.  For the first equation, there is an obvious solution
\begin{equation}
    \lambda=\frac{\kappa_2}{\kappa_1}\quad\mbox{and}\quad \gamma=\frac{\kappa_3}{\kappa_1}.
    \label{trivialsol}
\end{equation}
Therefore, using this solution, we get
\begin{equation}
    \kappa_1\Delta_1+\kappa_2\Delta_2+\kappa_3\Delta_3=0.
\end{equation}

\noindent\textbf{Lemma}. Suppose $\mathcal{P}$ satisfies $\mathcal{E} = 0$. Then $\Delta_1+\lambda \Delta_2+\gamma \Delta_3=0$ identically in $\mathcal{P}$ if and only if $(\lambda,\gamma)$ satisfies the relation (\ref{kinematic}); otherwise it requires the additional constraint (\ref{additional}) on $\mathcal{P}$. In particular, $(\lambda,\gamma)=\left(\frac{\kappa_2}{\kappa_1},\frac{\kappa_3}{\kappa_1}\right)$ always satisfies (\ref{kinematic}).

The constants $K_{12}=\kappa_1\kappa_2(\kappa_1-\kappa_2),$ $K_{13}=\kappa_1\kappa_3(\kappa_3-\kappa_1)$ and $K_{23}=\kappa_2\kappa_3(\kappa_2-\kappa_3)$ involved in Eq.(\ref{kinematic}) also have the interesting property :
\begin{equation}
    K_{23}+K_{13}+K_{12}=(\kappa_1-\kappa_2)(\kappa_1-\kappa_3)(\kappa_2-\kappa_3)\neq 0
\end{equation}
is the determinant of a Vandermonde matrix that is different from zero since the wave numbers $\kappa$ are assumed to be all distinct. This shows that Eq.(\ref{kinematic}) does not admit $\gamma=\lambda=1$ as a solution, which is coherent with the solution (\ref{trivialsol}).

This last lemma explains, from the integrable structure of SUSY bilinear KdV-type equations, the fermionic linear relation between the odd variables $\xi_1,\xi_2$ and $\xi_3$ established by Carstea \cite{Carstea} in the construction  of multi-soliton solution of some SUSY integrable equations.

\section{Results and discussion}

In this section, we classify for polynomial $\mathcal{P}$ that satisfy the Hirota integrability conditions established in the previous section. For such polynomials, we have constructed an unconstrained three-soliton solution. We will show that some polynomial $\mathcal{P}$ satisfies the classical constraint (\ref{BosoTest}) but fails to satisfy the other constraints. However, among these polynomials and under the fermionic relation established in the Lemma (see also Carstea \cite{Carstea}), we will show that they are associated to SUSY KdV-type bilinear equations possessing a three-soliton solution under the fermionic constraint and on the dispersion manifold. 

 Suppose that $\mathcal{P}$ is a polynomial for which equation $\mathcal{E}=0$ or equivalently equation (\ref{BosoTest}) is solved. This allows one to identify, as in the classical case \cite{10.1007/BFb0113694,Hietarinta1}, polynomials for which the purely bosonic integrability condition is satisfied. Let us first consider polynomials of the type
\begin{equation}
    \mathcal{P}(X,T)=a X^5+b X^3+c X+T,\label{GenPol}
\end{equation}
where $a,b,c$ are real constants. This polynomial satisfy the equation $\mathcal{E}=0$ given in (\ref{BosoTest}). This result being coherent with Hietarinta's bosonic classification \cite{Hietarinta1}. So, this polynomial constitute a good starting point in the identification of polynomials satisfying all the constraints. We now turn to the three fermionic integrability condition (\ref{F1}), (\ref{F2}) and (\ref{F3}). We show that, for free fermionic parameters $\xi_1$, $\xi_2$ and $\xi_3$, these conditions impose $a=0$ for the polynomial $\mathcal{P}$. 
This indicates that the $N=1$ SUSY Sawada-Kotera-Ramani equation studied by Carstea in \cite{Carstea} is non-integrable in the Hirota sense. The polynomial reduces to 
\begin{equation}
    \mathcal{P}(X,T)=b X^3+c X+T.
\end{equation}
This shows an important difference between the classical and SUSY classification: only a subset of classical bilinear KdV-type equation passes the fermionic constraints. The above polynomial satisfies all eight integrability conditions. We thus obtain a class of SUSY KdV-type bilinear equations passing the three-soliton test without any constraints on the parameters involved in the soliton construction:
\begin{equation}
    \mathcal{S}(b\mathcal{D}_x^3+c\mathcal{D}_x+\mathcal{D}_t)(\tau\cdot\tau)=0.
\end{equation}
This last equation naturally includes the supersymmetric KdV equation (\ref{SKdV}). Indeed, for this equation, we take $b=1$ and $c=0$.  Other polynomials satisfy the eight integrability conditions and thus possess an unconstrained SUSY three-soliton solution. For example, the polynomial \cite{DJZ,Hietarinta1}
 \begin{equation}
     \mathcal{P}(X,T)=T(X^2+\sqrt{3}XT+T^2)
 \end{equation}
 satisfy the eight integrability conditions. This polynomial was part of Hietarinta's classification \cite{Hietarinta1} in the classical case. We thus prove that this polynomial leads to not only three soliton solution in the classical case but also in the SUSY case. 

Going back to the general polynomial $\mathcal{P}$ given in equation (\ref{GenPol}). As we have previously mentioned this polynomial verifies the classical bosonic condition $\mathcal{E}=0$ given in (\ref{BosoTest}). However, for $a\neq0$, this polynomial does not allow the verification of the other seven integrability conditions which relied on the freedom of the fermionic parameters $\xi_1$, $\xi_2$ and $\xi_3$. This suggest that these odd parameters are linearly linked together.

However, Carstea  \cite{Carstea,Carstea1}, discussed other type of soliton construction where the fermionic parameters $\xi_1$, $\xi_2$ and $\xi_3$ are linearly related by:
\begin{eqnarray}
    \xi_1=\frac{\kappa_1}{\kappa_2}\xi_2=\frac{\kappa_1}{\kappa_3}\xi_3.\label{FermRel}
\end{eqnarray}
In this particular case, the eight integrability conditions are drastically simplified. Indeed, we are left with the classical bosonic integrability condition $\mathcal{E}=0$ given in equation (\ref{BosoTest}) which we know polynomial $\mathcal{P}$ to trivially satisfy. The fermionic integrability condition (\ref{FermionicEq}) reduces to
\begin{equation}
   \kappa_1 \Delta_1+\kappa_2\Delta_2+\kappa_3\Delta_3=0,
\end{equation}
as the contribution in $\xi_1\xi_2\xi_3=0$ using $\xi_1^2=0$. This last result is a consequence of the Lemma of the previous section. So the fermionic relation (\ref{FermRel}) established by Carstea is not only due to computational aspects but also from the integrability structure of such equations. In this case, where the fermionic parameters are linked, the $\tau-$function associated to the three-super-soliton (\ref{tau3Rep}) reduces to:
\begin{equation}
    \tau=1+e^{\check{\eta}_1}+e^{\check{\eta}_2}+e^{\check{\eta}_3}+B_{12}e^{\check{\eta}_1+\check{\eta}_2}+B_{13}e^{\check{\eta}_1+\check{\eta}_3}+B_{23}e^{\check{\eta}_2+\check{\eta}_3}+B_{12}B_{13}B_{23}e^{\check{\eta}_1+\check{\eta}_2+\check{\eta}_3},\label{tauRe}
\end{equation}
where $\check{\eta}_p=\eta_p+\theta\xi_p$ for $p=1,2,3$ and $B_{pq}$ are given in (\ref{BPQ}). This shows that when the fermionic parameters are linearly dependent, then the dressing method for the super-soliton solutions is unnecessary. Otherwise, the dressing method is mandatory to obtain exact fermionically unconstrained soliton solutions. We thus adopt the SUSY analogue of Hirota integrability:
\begin{enumerate}
    \item[$\bullet$] Strong Hirota integrability: \\
    We say that a SUSY KdV-type bilinear equation (\ref{BiliSKdV}) is strongly Hirota integrable if and only if the eight integrability conditions are satisfied without any constraints on the bosonic and fermionic parameters apart from the dispersion relations. In this case, the general three-super-soliton $\tau-$function is given by Eq.(\ref{tau3Rep}).
    \item[$\bullet$] Weak Hirota integrability:\\
    We say that a SUSY KdV-type bilinear equation is weakly Hirota integrable if the equation possesses a three-super-soliton $\tau-$function given by Eq.(\ref{tauRe}) under the fermionic relation (\ref{FermRel}).
\end{enumerate}
For example, this shows that the SUSY KdV equation (\ref{SKdV}) is strongly Hirota integrable, while the SUSY Sawada-Kotera-Ramani equation is weakly Hirota integrable. Note that strong Hirota integrability implies weak Hirota integrability, but not the other way around. 

Other polynomials relevant to Hietarinta's classification of bilinear KdV-type equations \cite{Hietarinta1} requires the fermionic relation to pass the SUSY analog of the three-soliton integrability test showing weak Hirota integrability. We mention, for example, the following polynomials:
\begin{equation}
    \mathcal{P}(X,T)=X^3-T^3+a X+ bT\quad \mbox{and}\quad\mathcal{P}(X,T)=X^2T-X-T.
\end{equation}
for arbitrary constants $a$ and $b$. The latter polynomial corresponds to the SUSY extension of the Hirota-Satsuma shallow-water-wave equation.

It is important to stress that our analysis can also be extended to larger spacial dimensions. Indeed, in \cite{Delisle}, the authors proposed a supersymmetric extension of the Boiti-Leon-Manna-Pempinelli (BLMP) equation. The bilinear form given, in the paper, is
\begin{equation}
    \mathcal{S}(\mathcal{D}_t+\mathcal{D}_y^3)(\tau\cdot\tau)=0.
\end{equation}
In the above formalism, the only difference is that $\eta=\kappa x+p y+\omega t+\varphi$. Taking $\mathcal{P}(Y,T)=Y^3+T$, we show that all eight Hirota integrability conditions are satisfied. This equation passes the SUSY three-soliton test. In this case, we have
\begin{equation}
    B_{mn}=\frac{(\kappa_m-\kappa_n)(p_m-p_n)}{(\kappa_m+\kappa_n)(p_m-p_n)}
\end{equation}
In \cite{Delisle}, the authors showed that an unconstrained two-soliton solution existed, but didn't proposed an unconstrained three super-soliton solution. The present formalism shows that the SUSY BLMP equation does indeed possess a free three-soliton solutions passing the SUSY Hirota's three soliton condition.

\section{Conclusion}

In this paper, we have presented supersymmetric bilinear equations of KdV type possessing unconstrained three super-soliton solution.
Our results provide a supersymmetric extension of Hietarinta's classical three-soliton classification by identifying the complete algebraic compatibility conditions governing unconstrained three-super-soliton solutions. In this sense, the present work establishes a unified framework for studying Hirota integrability in the presence of fermionic degrees of freedom. We have established fermionic and bosonic constraints for which such solutions exist. The purely bosonic constraint allows one to retrieve Hietarinta's classification of KdV-type bilinear equation passing the three-soliton test and showed that only a subset of this classification passes the three super-soliton integrability condition. For the other Hietarinta bilinear equations, we show that a linear relation must be established between the fermionic parameters involved in the soliton construction. This relation was found from a computational point of view by Carstea, here we show that this constraint is actually structural and depend on a constraint on the wave numbers of each solitary wave forming the three super-soliton solution. To make this distinction, we have introduce the concept of strong and weak Hirota integrability in the SUSY framework.

Furthermore, we have showed that susy KdV-type bilinear equation always possesses a two super-soliton solution and that the nonlinear coupling constant is dependent on the form of the polynomial $\mathcal{P}$ but also relies on an additional bosonic correction in $\xi_1\xi_2$.


Several natural extensions of this work remain open. Can the present classification be generalized to supersymmetric mKdV-, Sine-Gordon- and complex-type bilinear equations, where different algebraic structures are expected to emerge ? Another important direction is the construction of more general $\tau$-functions capable of relaxing the linear fermionic constraints, thereby extending the unrestricted supersymmetric classification established here.

\section*{Data Availability}
No data was generated or analyzed in this work.

\section*{Author Contributions}
Conceptualization: [Delisle, Jaouadi];
Formal analysis: [Delisle,Jaouadi];
Investigation: [Delisle,Jaoaudi];
Methodology: [Delisle,Jaouadi];
Software: [Delisle];
Validation: [Delisle,Jaouadi];
Visualization: [Delisle, Jaouadi];
Writing - original draft: [Delisle,Jaouadi]

\section*{Funding}
The authors did not receive support from any organization for the submitted work. The authors declare they have no financial interests.



\nocite{*}

\bibliography{apssamp}

\providecommand{\noopsort}[1]{}\providecommand{\singleletter}[1]{#1}%
\begin{thebibliography}{47}%
\makeatletter
\providecommand \@ifxundefined [1]{%
 \@ifx{#1\undefined}
}%
\providecommand \@ifnum [1]{%
 \ifnum #1\expandafter \@firstoftwo
 \else \expandafter \@secondoftwo
 \fi
}%
\providecommand \@ifx [1]{%
 \ifx #1\expandafter \@firstoftwo
 \else \expandafter \@secondoftwo
 \fi
}%
\providecommand \natexlab [1]{#1}%
\providecommand \enquote  [1]{``#1''}%
\providecommand \bibnamefont  [1]{#1}%
\providecommand \bibfnamefont [1]{#1}%
\providecommand \citenamefont [1]{#1}%
\providecommand \href@noop [0]{\@secondoftwo}%
\providecommand \href [0]{\begingroup \@sanitize@url \@href}%
\providecommand \@href[1]{\@@startlink{#1}\@@href}%
\providecommand \@@href[1]{\endgroup#1\@@endlink}%
\providecommand \@sanitize@url [0]{\catcode `\\12\catcode `\$12\catcode `\&12\catcode `\#12\catcode `\^12\catcode `\_12\catcode `\%12\relax}%
\providecommand \@@startlink[1]{}%
\providecommand \@@endlink[0]{}%
\providecommand \url  [0]{\begingroup\@sanitize@url \@url }%
\providecommand \@url [1]{\endgroup\@href {#1}{\urlprefix }}%
\providecommand \urlprefix  [0]{URL }%
\providecommand \Eprint [0]{\href }%
\providecommand \doibase [0]{https://doi.org/}%
\providecommand \selectlanguage [0]{\@gobble}%
\providecommand \bibinfo  [0]{\@secondoftwo}%
\providecommand \bibfield  [0]{\@secondoftwo}%
\providecommand \translation [1]{[#1]}%
\providecommand \BibitemOpen [0]{}%
\providecommand \bibitemStop [0]{}%
\providecommand \bibitemNoStop [0]{.\EOS\space}%
\providecommand \EOS [0]{\spacefactor3000\relax}%
\providecommand \BibitemShut  [1]{\csname bibitem#1\endcsname}%
\let\auto@bib@innerbib\@empty
\bibitem [{\citenamefont {Ablowitz}\ and\ \citenamefont {Segur}(1981)}]{Ablowitz}%
  \BibitemOpen
  \bibfield  {author} {\bibinfo {author} {\bibfnamefont {M.~J.}\ \bibnamefont {Ablowitz}}\ and\ \bibinfo {author} {\bibfnamefont {H.}~\bibnamefont {Segur}},\ }\href@noop {} {\emph {\bibinfo {title} {Solitons and the Inverse Scattering Transform}}}\ (\bibinfo  {publisher} {SIAM},\ \bibinfo {year} {1981})\BibitemShut {NoStop}%
\bibitem [{\citenamefont {Rogers}\ and\ \citenamefont {Schief}(2002)}]{Rogers}%
  \BibitemOpen
  \bibfield  {author} {\bibinfo {author} {\bibfnamefont {C.}~\bibnamefont {Rogers}}\ and\ \bibinfo {author} {\bibfnamefont {W.~K.}\ \bibnamefont {Schief}},\ }\href@noop {} {\emph {\bibinfo {title} {Bäcklund and {D}arboux transformations: geometry and modern applications in soliton theory}}}\ (\bibinfo  {publisher} {Cambridge University Press},\ \bibinfo {year} {2002})\BibitemShut {NoStop}%
\bibitem [{\citenamefont {Hirota}(2004)}]{Hirota}%
  \BibitemOpen
  \bibfield  {author} {\bibinfo {author} {\bibfnamefont {R.}~\bibnamefont {Hirota}},\ }\href@noop {} {\emph {\bibinfo {title} {The Direct Method in Soliton Theory}}}\ (\bibinfo  {publisher} {Cambridge University Press},\ \bibinfo {year} {2004})\BibitemShut {NoStop}%
\bibitem [{\citenamefont {Zhang}(2026)}]{DJZ}%
  \BibitemOpen
  \bibfield  {author} {\bibinfo {author} {\bibfnamefont {D.-J.}\ \bibnamefont {Zhang}},\ }\bibfield  {title} {\bibinfo {title} {Integrability and transformations in the bilinear method: {A}n introduction},\ }\bibfield  {journal} {\bibinfo  {journal} {Open Communications in Nonlinear Mathematical Physics, Special Issue in honour of Jarmo Hietarinta}\ }\textbf {\bibinfo {volume} {ocnmp: 18521}},\ \href {https://doi.org/10.46298/ocnmp.18521} {10.46298/ocnmp.18521} (\bibinfo {year} {2026})\BibitemShut {NoStop}%
\bibitem [{\citenamefont {Zhang}\ \emph {et~al.}(2025)\citenamefont {Zhang}, \citenamefont {Liu},\ and\ \citenamefont {Zhang}}]{zhang-liu-zhang}%
  \BibitemOpen
  \bibfield  {author} {\bibinfo {author} {\bibfnamefont {X.}~\bibnamefont {Zhang}}, \bibinfo {author} {\bibfnamefont {J.}~\bibnamefont {Liu}},\ and\ \bibinfo {author} {\bibfnamefont {D.-J.}\ \bibnamefont {Zhang}},\ }\bibfield  {title} {\bibinfo {title} {Nonlinearization of the {K}d{V}-type and m{K}d{V}-type bilinear equations},\ }\href@noop {} {\bibfield  {journal} {\bibinfo  {journal} {Commun. Theor. Phys.}\ }\textbf {\bibinfo {volume} {77}},\ \bibinfo {pages} {115006} (\bibinfo {year} {2025})}\BibitemShut {NoStop}%
\bibitem [{\citenamefont {Kumar}\ and\ \citenamefont {Mohan}(2022)}]{kumar-mohan-hbf}%
  \BibitemOpen
  \bibfield  {author} {\bibinfo {author} {\bibfnamefont {S.}~\bibnamefont {Kumar}}\ and\ \bibinfo {author} {\bibfnamefont {B.}~\bibnamefont {Mohan}},\ }\bibfield  {title} {\bibinfo {title} {A novel and efficient method for obtaining {H}irota's bilinear form for the nonlinear evolution equation in $(n+1)$ dimensions},\ }\href@noop {} {\bibfield  {journal} {\bibinfo  {journal} {Partial Differ.\ Eq.\ Appl.\ Math.}\ }\textbf {\bibinfo {volume} {5}},\ \bibinfo {pages} {100274} (\bibinfo {year} {2022})}\BibitemShut {NoStop}%
\bibitem [{\citenamefont {Razzaq}\ and\ \citenamefont {Zafar}(2025)}]{razzaq-zafar}%
  \BibitemOpen
  \bibfield  {author} {\bibinfo {author} {\bibfnamefont {W.}~\bibnamefont {Razzaq}}\ and\ \bibinfo {author} {\bibfnamefont {A.}~\bibnamefont {Zafar}},\ }\bibfield  {title} {\bibinfo {title} {Machine learning-enhanced soliton solutions for the {L}onngren-wave equation: an integration of {P}ainlev\'e analysis and {H}irota bilinear method},\ }\href@noop {} {\bibfield  {journal} {\bibinfo  {journal} {Rend.\ Fis.\ Acc.\ Lincei}\ }\textbf {\bibinfo {volume} {36}},\ \bibinfo {pages} {917} (\bibinfo {year} {2025})}\BibitemShut {NoStop}%
\bibitem [{\citenamefont {Ye}\ \emph {et~al.}(2011)\citenamefont {Ye}, \citenamefont {Wang}, \citenamefont {Chang},\ and\ \citenamefont {He}}]{ye-wang-chang-he}%
  \BibitemOpen
  \bibfield  {author} {\bibinfo {author} {\bibfnamefont {Y.}~\bibnamefont {Ye}}, \bibinfo {author} {\bibfnamefont {L.}~\bibnamefont {Wang}}, \bibinfo {author} {\bibfnamefont {Z.}~\bibnamefont {Chang}},\ and\ \bibinfo {author} {\bibfnamefont {J.}~\bibnamefont {He}},\ }\bibfield  {title} {\bibinfo {title} {An efficient algorithm of logarithmic transformation to {H}irota bilinear form of {K}d{V}-type bilinear equation},\ }\href@noop {} {\bibfield  {journal} {\bibinfo  {journal} {Applied Mathematics and Computation}\ }\textbf {\bibinfo {volume} {218}},\ \bibinfo {pages} {2200} (\bibinfo {year} {2011})}\BibitemShut {NoStop}%
\bibitem [{\citenamefont {Célanie}\ \emph {et~al.}(2024)\citenamefont {Célanie}, \citenamefont {Delisle},\ and\ \citenamefont {Jaouadi}}]{celanie}%
  \BibitemOpen
  \bibfield  {author} {\bibinfo {author} {\bibfnamefont {E.}~\bibnamefont {Célanie}}, \bibinfo {author} {\bibfnamefont {L.}~\bibnamefont {Delisle}},\ and\ \bibinfo {author} {\bibfnamefont {A.}~\bibnamefont {Jaouadi}},\ }\bibfield  {title} {\bibinfo {title} {Optically tuned soliton dynamics in {B}ose-{E}instein condensates within dark traps},\ }\href@noop {} {\bibfield  {journal} {\bibinfo  {journal} {J. Phys. A : Math. Theor.}\ }\textbf {\bibinfo {volume} {57}},\ \bibinfo {pages} {485701} (\bibinfo {year} {2024})}\BibitemShut {NoStop}%
\bibitem [{\citenamefont {Delisle}\ and\ \citenamefont {Jaouadi}(2025{\natexlab{a}})}]{delisle-symmetry}%
  \BibitemOpen
  \bibfield  {author} {\bibinfo {author} {\bibfnamefont {L.}~\bibnamefont {Delisle}}\ and\ \bibinfo {author} {\bibfnamefont {A.}~\bibnamefont {Jaouadi}},\ }\bibfield  {title} {\bibinfo {title} {Symmetry-driven multi-soliton dynamics in {B}ose-{E}instein condensates in reduced dimensions},\ }\href@noop {} {\bibfield  {journal} {\bibinfo  {journal} {Symmetry}\ }\textbf {\bibinfo {volume} {17(4)}},\ \bibinfo {pages} {582} (\bibinfo {year} {2025}{\natexlab{a}})}\BibitemShut {NoStop}%
\bibitem [{\citenamefont {Delisle}\ and\ \citenamefont {Jaouadi}(2025{\natexlab{b}})}]{10.1116/5.0304480}%
  \BibitemOpen
  \bibfield  {author} {\bibinfo {author} {\bibfnamefont {L.}~\bibnamefont {Delisle}}\ and\ \bibinfo {author} {\bibfnamefont {A.}~\bibnamefont {Jaouadi}},\ }\bibfield  {title} {\bibinfo {title} {Analytical control of multidark soliton collisions and dynamics with tunable properties},\ }\href@noop {} {\bibfield  {journal} {\bibinfo  {journal} {AVS Quantum Science}\ }\textbf {\bibinfo {volume} {7}},\ \bibinfo {pages} {041402} (\bibinfo {year} {2025}{\natexlab{b}})}\BibitemShut {NoStop}%
\bibitem [{\citenamefont {Delisle}\ \emph {et~al.}(2026)\citenamefont {Delisle}, \citenamefont {Hi},\ and\ \citenamefont {Jaouadi}}]{Delisle-finance}%
  \BibitemOpen
  \bibfield  {author} {\bibinfo {author} {\bibfnamefont {L.}~\bibnamefont {Delisle}}, \bibinfo {author} {\bibfnamefont {D.~P.}\ \bibnamefont {Hi}},\ and\ \bibinfo {author} {\bibfnamefont {A.}~\bibnamefont {Jaouadi}},\ }\bibfield  {title} {\bibinfo {title} {Bright solitons in a regulatory {I}vancevic model ({RIM}): analytical and numerical insights into price-wave dynamics},\ }\href@noop {} {\bibfield  {journal} {\bibinfo  {journal} {Nonlinear Dynamics}\ }\textbf {\bibinfo {volume} {114}},\ \bibinfo {pages} {329} (\bibinfo {year} {2026})}\BibitemShut {NoStop}%
\bibitem [{\citenamefont {Delisle}\ and\ \citenamefont {Jaouadi}(2026)}]{delisle-jaouadi}%
  \BibitemOpen
  \bibfield  {author} {\bibinfo {author} {\bibfnamefont {L.}~\bibnamefont {Delisle}}\ and\ \bibinfo {author} {\bibfnamefont {A.}~\bibnamefont {Jaouadi}},\ }\bibfield  {title} {\bibinfo {title} {A vector bilinear framework for soliton dynamics in coupled modified {K}d{V} systems},\ }\bibfield  {journal} {\bibinfo  {journal} {Open Communications in Nonlinear Mathematical Physics, Special Issue in honour of Jarmo Hietarinta}\ }\textbf {\bibinfo {volume} {ocnmp:18025}},\ \href {https://doi.org/10.46298/ocnmp.18025} {10.46298/ocnmp.18025} (\bibinfo {year} {2026})\BibitemShut {NoStop}%
\bibitem [{\citenamefont {Foucher}\ \emph {et~al.}(2026)\citenamefont {Foucher}, \citenamefont {Delisle},\ and\ \citenamefont {Jaouadi}}]{foucher}%
  \BibitemOpen
  \bibfield  {author} {\bibinfo {author} {\bibfnamefont {M.}~\bibnamefont {Foucher}}, \bibinfo {author} {\bibfnamefont {L.}~\bibnamefont {Delisle}},\ and\ \bibinfo {author} {\bibfnamefont {A.}~\bibnamefont {Jaouadi}},\ }\bibfield  {title} {\bibinfo {title} {Vector representation of exact soliton dynamics in multi-component nonlinear {S}chrödinger systems},\ }\href@noop {} {\bibfield  {journal} {\bibinfo  {journal} {arXiv: 2606.27569}\ } (\bibinfo {year} {2026})}\BibitemShut {NoStop}%
\bibitem [{\citenamefont {Hietarinta}(1997)}]{10.1007/BFb0113694}%
  \BibitemOpen
  \bibfield  {author} {\bibinfo {author} {\bibfnamefont {J.}~\bibnamefont {Hietarinta}},\ }\bibfield  {title} {\bibinfo {title} {Introduction to the {H}irota bilinear method},\ }in\ \href@noop {} {\emph {\bibinfo {booktitle} {Integrability of Nonlinear Systems}}},\ \bibinfo {editor} {edited by\ \bibinfo {editor} {\bibfnamefont {Y.}~\bibnamefont {Kosmann-Schwarzbach}}, \bibinfo {editor} {\bibfnamefont {B.}~\bibnamefont {Grammaticos}},\ and\ \bibinfo {editor} {\bibfnamefont {K.~M.}\ \bibnamefont {Tamizhmani}}}\ (\bibinfo  {publisher} {Springer Berlin Heidelberg},\ \bibinfo {address} {Berlin, Heidelberg},\ \bibinfo {year} {1997})\ pp.\ \bibinfo {pages} {95--103}\BibitemShut {NoStop}%
\bibitem [{\citenamefont {Hietarinta}(1987{\natexlab{a}})}]{Hietarinta1}%
  \BibitemOpen
  \bibfield  {author} {\bibinfo {author} {\bibfnamefont {J.}~\bibnamefont {Hietarinta}},\ }\bibfield  {title} {\bibinfo {title} {A search for bilinear equations passing {H}irota's three soliton condition. {I}. {K}d{V}-type bilinear equations},\ }\href@noop {} {\bibfield  {journal} {\bibinfo  {journal} {J.\ Math.\ Phys.}\ }\textbf {\bibinfo {volume} {28}},\ \bibinfo {pages} {1732} (\bibinfo {year} {1987}{\natexlab{a}})}\BibitemShut {NoStop}%
\bibitem [{\citenamefont {Hietarinta}(1987{\natexlab{b}})}]{Hietarinta2}%
  \BibitemOpen
  \bibfield  {author} {\bibinfo {author} {\bibfnamefont {J.}~\bibnamefont {Hietarinta}},\ }\bibfield  {title} {\bibinfo {title} {A search for bilinear equations passing {H}irota's three soliton condition. {II}. m{K}d{V}-type bilinear equations},\ }\href@noop {} {\bibfield  {journal} {\bibinfo  {journal} {J.\ Math.\ Phys.}\ }\textbf {\bibinfo {volume} {28}},\ \bibinfo {pages} {2094} (\bibinfo {year} {1987}{\natexlab{b}})}\BibitemShut {NoStop}%
\bibitem [{\citenamefont {Hietarinta}(1987{\natexlab{c}})}]{Hietarinta3}%
  \BibitemOpen
  \bibfield  {author} {\bibinfo {author} {\bibfnamefont {J.}~\bibnamefont {Hietarinta}},\ }\bibfield  {title} {\bibinfo {title} {A search for bilinear equations passing {H}irota's three soliton condition. {III}. {S}ine-{G}ordon-type bilinear equations},\ }\href@noop {} {\bibfield  {journal} {\bibinfo  {journal} {J.\ Math.\ Phys.}\ }\textbf {\bibinfo {volume} {28}},\ \bibinfo {pages} {2586} (\bibinfo {year} {1987}{\natexlab{c}})}\BibitemShut {NoStop}%
\bibitem [{\citenamefont {Hietarinta}(1988)}]{Hietarinta4}%
  \BibitemOpen
  \bibfield  {author} {\bibinfo {author} {\bibfnamefont {J.}~\bibnamefont {Hietarinta}},\ }\bibfield  {title} {\bibinfo {title} {A search for bilinear equations passing {H}irota's three soliton condition. {IV}. {C}omplex bilinear equations},\ }\href@noop {} {\bibfield  {journal} {\bibinfo  {journal} {J.\ Math.\ Phys.}\ }\textbf {\bibinfo {volume} {29}},\ \bibinfo {pages} {628} (\bibinfo {year} {1988})}\BibitemShut {NoStop}%
\bibitem [{\citenamefont {Popowicz}(1993)}]{popowicz-lax}%
  \BibitemOpen
  \bibfield  {author} {\bibinfo {author} {\bibfnamefont {Z.}~\bibnamefont {Popowicz}},\ }\bibfield  {title} {\bibinfo {title} {The {L}ax formulation of the {\rm ``new''} {N}$=2$ {SUSY} {K}d{V} equation},\ }\href@noop {} {\bibfield  {journal} {\bibinfo  {journal} {Phys.\ Lett.\ A}\ }\textbf {\bibinfo {volume} {174}},\ \bibinfo {pages} {411} (\bibinfo {year} {1993})}\BibitemShut {NoStop}%
\bibitem [{\citenamefont {Krivonos}\ \emph {et~al.}(1998)\citenamefont {Krivonos}, \citenamefont {Pashnev},\ and\ \citenamefont {Popowicz}}]{krivonos-pashnev-popowicz}%
  \BibitemOpen
  \bibfield  {author} {\bibinfo {author} {\bibfnamefont {S.}~\bibnamefont {Krivonos}}, \bibinfo {author} {\bibfnamefont {A.}~\bibnamefont {Pashnev}},\ and\ \bibinfo {author} {\bibfnamefont {Z.}~\bibnamefont {Popowicz}},\ }\bibfield  {title} {\bibinfo {title} {Lax pairs for {N}$=2,3$ supersymmetric {K}d{V} equations and their extensions},\ }\href@noop {} {\bibfield  {journal} {\bibinfo  {journal} {Mod.\ Phys.\ Lett.\ A}\ }\textbf {\bibinfo {volume} {13}},\ \bibinfo {pages} {1435} (\bibinfo {year} {1998})}\BibitemShut {NoStop}%
\bibitem [{\citenamefont {Oevel}\ and\ \citenamefont {Popowicz}(1991)}]{Oevel}%
  \BibitemOpen
  \bibfield  {author} {\bibinfo {author} {\bibfnamefont {W.}~\bibnamefont {Oevel}}\ and\ \bibinfo {author} {\bibfnamefont {Z.}~\bibnamefont {Popowicz}},\ }\bibfield  {title} {\bibinfo {title} {The {B}i-{H}amiltonian structure of fully supersymmetric {K}orteweg-de {V}ries systems},\ }\href@noop {} {\bibfield  {journal} {\bibinfo  {journal} {Commun.\ Math.\ Phys.}\ }\textbf {\bibinfo {volume} {139}},\ \bibinfo {pages} {441} (\bibinfo {year} {1991})}\BibitemShut {NoStop}%
\bibitem [{\citenamefont {Andrea}\ \emph {et~al.}(2005)\citenamefont {Andrea}, \citenamefont {Restuccia},\ and\ \citenamefont {Sotomayor}}]{andrea-restuccia-sotomayor}%
  \BibitemOpen
  \bibfield  {author} {\bibinfo {author} {\bibfnamefont {S.}~\bibnamefont {Andrea}}, \bibinfo {author} {\bibfnamefont {A.}~\bibnamefont {Restuccia}},\ and\ \bibinfo {author} {\bibfnamefont {A.}~\bibnamefont {Sotomayor}},\ }\bibfield  {title} {\bibinfo {title} {The {G}ardner category and non-local conservation laws for {N}$=1$ super {K}d{V}},\ }\href@noop {} {\bibfield  {journal} {\bibinfo  {journal} {J.\ Math.\ Phys.}\ }\textbf {\bibinfo {volume} {46}},\ \bibinfo {pages} {103517} (\bibinfo {year} {2005})}\BibitemShut {NoStop}%
\bibitem [{\citenamefont {Dargis}\ and\ \citenamefont {Mathieu}(1993)}]{dargis-mathieu}%
  \BibitemOpen
  \bibfield  {author} {\bibinfo {author} {\bibfnamefont {P.}~\bibnamefont {Dargis}}\ and\ \bibinfo {author} {\bibfnamefont {P.}~\bibnamefont {Mathieu}},\ }\bibfield  {title} {\bibinfo {title} {Nonlocal conservation laws in {N}$=1,2$ supersymmetric {K}d{V} equation},\ }\href@noop {} {\bibfield  {journal} {\bibinfo  {journal} {Phys.\ Lett.\ A}\ }\textbf {\bibinfo {volume} {176}},\ \bibinfo {pages} {67} (\bibinfo {year} {1993})}\BibitemShut {NoStop}%
\bibitem [{\citenamefont {Manin}\ and\ \citenamefont {Radul}(1985)}]{Manin}%
  \BibitemOpen
  \bibfield  {author} {\bibinfo {author} {\bibfnamefont {Y.~I.}\ \bibnamefont {Manin}}\ and\ \bibinfo {author} {\bibfnamefont {A.~O.}\ \bibnamefont {Radul}},\ }\bibfield  {title} {\bibinfo {title} {A supersymmetric extension of the {K}adomtsev-{P}etviashvili hierarchy},\ }\href@noop {} {\bibfield  {journal} {\bibinfo  {journal} {Commun.\ Math.\ Phys.}\ }\textbf {\bibinfo {volume} {98}},\ \bibinfo {pages} {65} (\bibinfo {year} {1985})}\BibitemShut {NoStop}%
\bibitem [{\citenamefont {Ivanov}\ and\ \citenamefont {Krivonos}(1997)}]{ivanov-krivonos}%
  \BibitemOpen
  \bibfield  {author} {\bibinfo {author} {\bibfnamefont {E.}~\bibnamefont {Ivanov}}\ and\ \bibinfo {author} {\bibfnamefont {S.}~\bibnamefont {Krivonos}},\ }\bibfield  {title} {\bibinfo {title} {New integrable extensions of {N}$=2$ {K}d{V} and {B}oussinesq hierarchies},\ }\href@noop {} {\bibfield  {journal} {\bibinfo  {journal} {Phys.\ Lett.\ A}\ }\textbf {\bibinfo {volume} {231}},\ \bibinfo {pages} {75} (\bibinfo {year} {1997})}\BibitemShut {NoStop}%
\bibitem [{\citenamefont {McArthur}\ and\ \citenamefont {Yung}(1993)}]{McArthur}%
  \BibitemOpen
  \bibfield  {author} {\bibinfo {author} {\bibfnamefont {I.~N.}\ \bibnamefont {McArthur}}\ and\ \bibinfo {author} {\bibfnamefont {C.~M.}\ \bibnamefont {Yung}},\ }\bibfield  {title} {\bibinfo {title} {Hirota bilinear form for the super-{K}d{V} hierarchy},\ }\href@noop {} {\bibfield  {journal} {\bibinfo  {journal} {Modern Physics Letters A}\ }\textbf {\bibinfo {volume} {8}},\ \bibinfo {pages} {1739} (\bibinfo {year} {1993})}\BibitemShut {NoStop}%
\bibitem [{\citenamefont {Figueroa-O'Farrill}\ \emph {et~al.}(1991)\citenamefont {Figueroa-O'Farrill}, \citenamefont {Ramos},\ and\ \citenamefont {Mas}}]{Figueroa}%
  \BibitemOpen
  \bibfield  {author} {\bibinfo {author} {\bibfnamefont {J.~M.}\ \bibnamefont {Figueroa-O'Farrill}}, \bibinfo {author} {\bibfnamefont {E.}~\bibnamefont {Ramos}},\ and\ \bibinfo {author} {\bibfnamefont {J.}~\bibnamefont {Mas}},\ }\bibfield  {title} {\bibinfo {title} {Integrability and bihalmitonian structure of the even order {SK}d{V} hierarchies},\ }\href@noop {} {\bibfield  {journal} {\bibinfo  {journal} {Reviews in Mathematical Physics}\ }\textbf {\bibinfo {volume} {3}},\ \bibinfo {pages} {479} (\bibinfo {year} {1991})}\BibitemShut {NoStop}%
\bibitem [{\citenamefont {Mathieu}(1988)}]{mathieu}%
  \BibitemOpen
  \bibfield  {author} {\bibinfo {author} {\bibfnamefont {P.}~\bibnamefont {Mathieu}},\ }\bibfield  {title} {\bibinfo {title} {Supersymmetric extension of the {K}orteweg-de {V}ries equation},\ }\href@noop {} {\bibfield  {journal} {\bibinfo  {journal} {J.\ Math.\ Phys.}\ }\textbf {\bibinfo {volume} {29}},\ \bibinfo {pages} {2499} (\bibinfo {year} {1988})}\BibitemShut {NoStop}%
\bibitem [{\citenamefont {Labelle}\ and\ \citenamefont {Mathieu}(1991)}]{labelle-mathieu}%
  \BibitemOpen
  \bibfield  {author} {\bibinfo {author} {\bibfnamefont {P.}~\bibnamefont {Labelle}}\ and\ \bibinfo {author} {\bibfnamefont {P.}~\bibnamefont {Mathieu}},\ }\bibfield  {title} {\bibinfo {title} {A new {N}$=2$ supersymmetric {K}orteweg-de {V}ries equation},\ }\href@noop {} {\bibfield  {journal} {\bibinfo  {journal} {J.\ Math.\ Phys.}\ }\textbf {\bibinfo {volume} {32}},\ \bibinfo {pages} {923} (\bibinfo {year} {1991})}\BibitemShut {NoStop}%
\bibitem [{\citenamefont {Carstea}(2000)}]{Carstea}%
  \BibitemOpen
  \bibfield  {author} {\bibinfo {author} {\bibfnamefont {A.~S.}\ \bibnamefont {Carstea}},\ }\bibfield  {title} {\bibinfo {title} {Extension of the bilinear formalism to supersymmetric {K}d{V}-type equations},\ }\href@noop {} {\bibfield  {journal} {\bibinfo  {journal} {Nonlinearity}\ }\textbf {\bibinfo {volume} {13}},\ \bibinfo {pages} {1645} (\bibinfo {year} {2000})}\BibitemShut {NoStop}%
\bibitem [{\citenamefont {A.~S.~Carstea}\ and\ \citenamefont {Grammaticos}(2001)}]{Carstea1}%
  \BibitemOpen
  \bibfield  {author} {\bibinfo {author} {\bibfnamefont {A.~R.}\ \bibnamefont {A.~S.~Carstea}}\ and\ \bibinfo {author} {\bibfnamefont {B.}~\bibnamefont {Grammaticos}},\ }\bibfield  {title} {\bibinfo {title} {Constructing the soliton solutions for the {N}$=1$ supersymmetric {K}d{V} hierarchy},\ }\href@noop {} {\bibfield  {journal} {\bibinfo  {journal} {Nonlinearity}\ }\textbf {\bibinfo {volume} {14}},\ \bibinfo {pages} {1419} (\bibinfo {year} {2001})}\BibitemShut {NoStop}%
\bibitem [{\citenamefont {Liu}\ and\ \citenamefont {Hu}(2005)}]{Liu}%
  \BibitemOpen
  \bibfield  {author} {\bibinfo {author} {\bibfnamefont {Q.~P.}\ \bibnamefont {Liu}}\ and\ \bibinfo {author} {\bibfnamefont {X.-B.}\ \bibnamefont {Hu}},\ }\bibfield  {title} {\bibinfo {title} {Bilinearization of {N}$=1$ supersymmetric {K}orteweg-de {V}ries equation revisited},\ }\href@noop {} {\bibfield  {journal} {\bibinfo  {journal} {J.\ Phys.\ A : Math.\ Gen.}\ }\textbf {\bibinfo {volume} {38}},\ \bibinfo {pages} {6371} (\bibinfo {year} {2005})}\BibitemShut {NoStop}%
\bibitem [{\citenamefont {Ghosh}\ and\ \citenamefont {Sarma}(2003)}]{Ghosh}%
  \BibitemOpen
  \bibfield  {author} {\bibinfo {author} {\bibfnamefont {S.}~\bibnamefont {Ghosh}}\ and\ \bibinfo {author} {\bibfnamefont {D.}~\bibnamefont {Sarma}},\ }\bibfield  {title} {\bibinfo {title} {Bilinearization of {N}$=1$ supersymmetric modified {K}d{V} equations},\ }\href@noop {} {\bibfield  {journal} {\bibinfo  {journal} {Nonlinearity}\ }\textbf {\bibinfo {volume} {16}},\ \bibinfo {pages} {411} (\bibinfo {year} {2003})}\BibitemShut {NoStop}%
\bibitem [{\citenamefont {Delisle}\ and\ \citenamefont {Mosaddeghi}(2013)}]{Delisle}%
  \BibitemOpen
  \bibfield  {author} {\bibinfo {author} {\bibfnamefont {L.}~\bibnamefont {Delisle}}\ and\ \bibinfo {author} {\bibfnamefont {M.}~\bibnamefont {Mosaddeghi}},\ }\bibfield  {title} {\bibinfo {title} {Classical and {SUSY} solutions of the {B}oiti-{L}eon-{M}anna-{P}empinelli equation},\ }\href@noop {} {\bibfield  {journal} {\bibinfo  {journal} {J.\ Phys.\ A : Math.\ Theor.}\ }\textbf {\bibinfo {volume} {46}},\ \bibinfo {pages} {115203} (\bibinfo {year} {2013})}\BibitemShut {NoStop}%
\bibitem [{\citenamefont {Delisle}(2023)}]{delisle-novel-hbf}%
  \BibitemOpen
  \bibfield  {author} {\bibinfo {author} {\bibfnamefont {L.}~\bibnamefont {Delisle}},\ }\bibfield  {title} {\bibinfo {title} {A novel {H}irota bilinear approach to {N}$=2$ supersymmetric equations},\ }\href@noop {} {\bibfield  {journal} {\bibinfo  {journal} {J. Phys. A: Math. Theor.}\ }\textbf {\bibinfo {volume} {56}},\ \bibinfo {pages} {455202} (\bibinfo {year} {2023})}\BibitemShut {NoStop}%
\bibitem [{\citenamefont {Ghosh}\ and\ \citenamefont {Sarma}(2001)}]{ghosh-sarma-2001}%
  \BibitemOpen
  \bibfield  {author} {\bibinfo {author} {\bibfnamefont {S.}~\bibnamefont {Ghosh}}\ and\ \bibinfo {author} {\bibfnamefont {D.}~\bibnamefont {Sarma}},\ }\bibfield  {title} {\bibinfo {title} {Soliton solutions for the {N}$=2$ supersymmetric {K}d{V} equation},\ }\href@noop {} {\bibfield  {journal} {\bibinfo  {journal} {Phys.\ Lett.\ B}\ }\textbf {\bibinfo {volume} {522}},\ \bibinfo {pages} {189} (\bibinfo {year} {2001})}\BibitemShut {NoStop}%
\bibitem [{\citenamefont {Delisle}(2017)}]{delisle-n2-hbf-2017}%
  \BibitemOpen
  \bibfield  {author} {\bibinfo {author} {\bibfnamefont {L.}~\bibnamefont {Delisle}},\ }\bibfield  {title} {\bibinfo {title} {A {N}$=2$ extension of the {H}irota bilinear formalism and the supersymmetric {K}d{V} equation},\ }\href@noop {} {\bibfield  {journal} {\bibinfo  {journal} {J.\ Math.\ Phys.}\ }\textbf {\bibinfo {volume} {58}},\ \bibinfo {pages} {013504} (\bibinfo {year} {2017})}\BibitemShut {NoStop}%
\bibitem [{\citenamefont {Liu}\ \emph {et~al.}(2005)\citenamefont {Liu}, \citenamefont {Hu},\ and\ \citenamefont {Zhang}}]{LiuHuZhang}%
  \BibitemOpen
  \bibfield  {author} {\bibinfo {author} {\bibfnamefont {Q.~P.}\ \bibnamefont {Liu}}, \bibinfo {author} {\bibfnamefont {X.-B.}\ \bibnamefont {Hu}},\ and\ \bibinfo {author} {\bibfnamefont {M.-X.}\ \bibnamefont {Zhang}},\ }\bibfield  {title} {\bibinfo {title} {Supersymmetric modified {K}orteweg-de {V}ries equation: bilinear approach},\ }\href@noop {} {\bibfield  {journal} {\bibinfo  {journal} {Nonlinearity}\ }\textbf {\bibinfo {volume} {18}},\ \bibinfo {pages} {1597} (\bibinfo {year} {2005})}\BibitemShut {NoStop}%
\bibitem [{\citenamefont {Liu}\ and\ \citenamefont {{M. Ma\~nas}}(1997{\natexlab{a}})}]{liu-manas-darboux}%
  \BibitemOpen
  \bibfield  {author} {\bibinfo {author} {\bibfnamefont {Q.~P.}\ \bibnamefont {Liu}}\ and\ \bibinfo {author} {\bibnamefont {{M. Ma\~nas}}},\ }\bibfield  {title} {\bibinfo {title} {Darboux transformation for the {M}anin-{R}adul supersymmetric {K}d{V} equation},\ }\href@noop {} {\bibfield  {journal} {\bibinfo  {journal} {Phys.\ Lett.\ B}\ }\textbf {\bibinfo {volume} {394}},\ \bibinfo {pages} {337} (\bibinfo {year} {1997}{\natexlab{a}})}\BibitemShut {NoStop}%
\bibitem [{\citenamefont {Liu}\ and\ \citenamefont {{M. Ma\~nas}}(1997{\natexlab{b}})}]{liu-manas-crum}%
  \BibitemOpen
  \bibfield  {author} {\bibinfo {author} {\bibfnamefont {Q.~P.}\ \bibnamefont {Liu}}\ and\ \bibinfo {author} {\bibnamefont {{M. Ma\~nas}}},\ }\bibfield  {title} {\bibinfo {title} {Crum transformation and wronskian type solutions for supersymmetric {K}d{V} equation},\ }\href@noop {} {\bibfield  {journal} {\bibinfo  {journal} {Phys.\ Lett.\ B}\ }\textbf {\bibinfo {volume} {396}},\ \bibinfo {pages} {133} (\bibinfo {year} {1997}{\natexlab{b}})}\BibitemShut {NoStop}%
\bibitem [{\citenamefont {Liu}\ and\ \citenamefont {{M. Ma\~nas}}(1998)}]{liu-manas-pfaffian}%
  \BibitemOpen
  \bibfield  {author} {\bibinfo {author} {\bibfnamefont {Q.~P.}\ \bibnamefont {Liu}}\ and\ \bibinfo {author} {\bibnamefont {{M. Ma\~nas}}},\ }\bibfield  {title} {\bibinfo {title} {Pfaffian solutions for the {M}anin-{R}adul-{M}athieu {SUSY} {K}d{V} and {SUSY} sine-{G}ordon equations},\ }\href@noop {} {\bibfield  {journal} {\bibinfo  {journal} {Phys.\ Lett.\ B}\ }\textbf {\bibinfo {volume} {436}},\ \bibinfo {pages} {306} (\bibinfo {year} {1998})}\BibitemShut {NoStop}%
\bibitem [{\citenamefont {Liu}(1995)}]{liu-darboux-1995}%
  \BibitemOpen
  \bibfield  {author} {\bibinfo {author} {\bibfnamefont {Q.~P.}\ \bibnamefont {Liu}},\ }\bibfield  {title} {\bibinfo {title} {Darboux transformations for supersymmetric {K}orteweg-de {V}ries equations},\ }\href@noop {} {\bibfield  {journal} {\bibinfo  {journal} {Lett.\ Math.\ Phys.}\ }\textbf {\bibinfo {volume} {35}},\ \bibinfo {pages} {115} (\bibinfo {year} {1995})}\BibitemShut {NoStop}%
\bibitem [{\citenamefont {Liu}\ and\ \citenamefont {Xie}(2004)}]{liu-xie}%
  \BibitemOpen
  \bibfield  {author} {\bibinfo {author} {\bibfnamefont {Q.~P.}\ \bibnamefont {Liu}}\ and\ \bibinfo {author} {\bibfnamefont {Y.~F.}\ \bibnamefont {Xie}},\ }\bibfield  {title} {\bibinfo {title} {Nonlinear superposition formula for {N}$=1$ supersymmetric {K}d{V} equation},\ }\href@noop {} {\bibfield  {journal} {\bibinfo  {journal} {Phys.\ Lett.\ A}\ }\textbf {\bibinfo {volume} {325}},\ \bibinfo {pages} {139} (\bibinfo {year} {2004})}\BibitemShut {NoStop}%
\bibitem [{\citenamefont {Li}\ and\ \citenamefont {Nimmo}(2010)}]{li-nimmo}%
  \BibitemOpen
  \bibfield  {author} {\bibinfo {author} {\bibfnamefont {C.~X.}\ \bibnamefont {Li}}\ and\ \bibinfo {author} {\bibfnamefont {J.~J.~C.}\ \bibnamefont {Nimmo}},\ }\bibfield  {title} {\bibinfo {title} {Darboux transformations for a twisted derivation and quasideterminant solutions to the super {K}d{V} equation},\ }\href@noop {} {\bibfield  {journal} {\bibinfo  {journal} {Proc.\ R.\ Soc.\ A}\ }\textbf {\bibinfo {volume} {466}},\ \bibinfo {pages} {2471} (\bibinfo {year} {2010})}\BibitemShut {NoStop}%
\bibitem [{\citenamefont {Fan}\ and\ \citenamefont {Hon}(2012)}]{fan-hon-2012}%
  \BibitemOpen
  \bibfield  {author} {\bibinfo {author} {\bibfnamefont {E.}~\bibnamefont {Fan}}\ and\ \bibinfo {author} {\bibfnamefont {Y.~C.}\ \bibnamefont {Hon}},\ }\bibfield  {title} {\bibinfo {title} {Super extension of {B}ell polynomials with applications to supersymmetric equations},\ }\href@noop {} {\bibfield  {journal} {\bibinfo  {journal} {J.\ Math.\ Phys.}\ }\textbf {\bibinfo {volume} {53}},\ \bibinfo {pages} {013503} (\bibinfo {year} {2012})}\BibitemShut {NoStop}%
\bibitem [{\citenamefont {Cornwell}(1989)}]{Cornwell:1989bx}%
  \BibitemOpen
  \bibfield  {author} {\bibinfo {author} {\bibfnamefont {J.~F.}\ \bibnamefont {Cornwell}},\ }\href@noop {} {\emph {\bibinfo {title} {{Group Theory in Physics. Volume III: Supersymmetries and Infinite-Dimensional Algebras}}}},\ \bibinfo {series} {Techniques of Physics}, Vol.~\bibinfo {volume} {10}\ (\bibinfo  {publisher} {Academic Press},\ \bibinfo {address} {London},\ \bibinfo {year} {1989})\BibitemShut {NoStop}%
\end{thebibliography}%

\end{document}